\documentclass{nature}

\usepackage{amsmath}
\usepackage[pdftex]{graphicx}
\makeatletter
\let\saved@includegraphics\includegraphics
\AtBeginDocument{\let\includegraphics\saved@includegraphics}
\renewenvironment*{figure}{\@float{figure}}{\end@float}
\makeatother

\usepackage{dcolumn}  % Align table columns on decimal point
\usepackage{bm}       % bold math
\usepackage{dsfont}
\usepackage{color}
\usepackage{braket}

\newcounter{defcounter}
\usepackage{amsthm}

\title{ Disentangling Dynamical Quantum Coherences in the Fenna-Matthews-Olson Complex } 

\author{Hong-Guang Duan$^{1,2,3,*}$, Ajay Jha$^{1,4,*}$, Lipeng Chen$^{5,*}$, Vandana Tiwari$^{1,6}$, Richard J. Cogdell$^{7}$, Khuram Ashraf$^{7}$, Valentyn I. Prokhorenko$^{1}$, Michael Thorwart$^{2,3}$ \& R. J. Dwayne Miller$^{1,3,8}$} 

\begin{document} 

\maketitle

\begin{affiliations}
\item Max Planck Institute for the Structure and Dynamics of Matter, Luruper
Chaussee 149, 22761, Hamburg, Germany
\item I.\ Institut f\"ur Theoretische Physik,  Universit\"at Hamburg,
Jungiusstra{\ss}e 9, 20355 Hamburg, Germany
\item The Hamburg Center for Ultrafast Imaging, Luruper Chaussee 149, 22761
Hamburg, Germany
\item The Rosalind Franklin Institute, Rutherford Appleton Laboratory, Harwell Campus, Didcot, Oxfordshire, OX11 OFA, United Kingdom
\item Department of Chemistry, Technische Universit{\"a}t M{\"u}nchen, D-85747 Garching, Germany  
\item Department of Chemistry, Universit\"at Hamburg, Martin-Luther-King Platz 6, 20146 Hamburg, Germany
\item Institute of Molecular, Cell, and Systems Biology, College of Medical, Veterinary, and Life Science, University of Glasgow, Glasgow G12 8QQ, United Kingdom 
\item The Departments of Chemistry and Physics, University of Toronto, 80 St.
George Street, Toronto Canada M5S 3H6\\
$^*$These authors contributed equally to this work. \\
\centerline{\underline{\date{\bf \today}}}
\end{affiliations} 

\begin{abstract} 

In the primary step of natural light-harvesting, the energy of a solar photon is captured in antenna chlorophyll as a photoexcited electron-hole pair, or an exciton. Its efficient conversion to stored chemical potential occurs in the special pair reaction center, which has to be reached by down-hill ultrafast excited state energy transport. Key to this process is the degree of interaction between the chlorophyll chromophores that can lead to spatial delocalization and quantum coherence effects. The importance of quantum contributions to energy transport depends on the relative coupling between the chlorophylls in relation to the intensity of the fluctuations and reorganization dynamics of the surrounding protein matrix, or bath. The latter induce uncorrelated modulations of the site energies, resulting in quantum decoherence, and localization of the spatial extent of the exciton. %The key issue determining the significance of quantum coherence in this biological process is the degree of coupling between the chlorophyll molecules relative to the bath. 
%The bath coupling can be directly probed by the temperature dependence of the quantum decoherence dynamics as a means to systematically vary the frequency and amplitude of the surrounding bath motions. 
The current consensus is that under physiological conditions quantum decoherence occurs on the 10 fs time scale, and quantum coherence plays little role for the observed picosecond energy transfer dynamics. In this work, we reaffirm this from a different point of view by finding that the true onset of important electronic quantum coherence only occurs at extremely low temperatures of $\sim$20 K. We have directly determined the exciton coherence times using two-dimensional (2D) electronic spectroscopy of the Fenna-Matthew-Olson (FMO) complex over an extensive temperature range. At 20 K, we show that electronic coherences persist out to 200 fs (close to the antenna) and marginally up to 500 fs at the reaction-center side. The electronic coherence is found to decay markedly faster with modest increases in temperature to become irrelevant above 150 K. This temperature dependence also allows disentangling the previously reported long-lived beatings thought to be evidence for electronic quantum coherence contributions. We show that they result from mixing vibrational coherences in the electronic ground state. We also uncover the relevant electronic coherence between excited electronic states and examine the temperature-dependent non-Markovianity of the transfer dynamics to show that the bath involves uncorrelated motions even to low temperatures. The observed temperature dependence allows a clear separation of the fragile electronic coherence from the robust vibrational coherence. The specific details of the critical bath interaction are treated through a theoretical model based on measured bath parameters that reproduces the temperature dependent dynamics. By this, we provide a complete picture of the bath interaction which places these systems in the regime of strong bath coupling. We believe this main conclusion to be generically valid for light harvesting systems. This principle makes the systems robust against otherwise fragile quantum effects as evidenced by the strong temperature dependence. We conclude that nature explicitly exploits decoherence or dissipation in engineering site energies to yield downhill energy gradients to unerringly direct energy, even on the fastest time scales of biological processes. 

\end{abstract} 

%%%%%%%%%%%% introduction

The question how biological function emerges from the atomic constituents of matter has intrigued scientists since the early days of quantum mechanics \cite{Jordan,Book1}. Encouraged by the success of quantum theory to describe matter, its pioneers rapidly explored extending it to the world of chemistry and biology in the early 20th century. It was only in recent times when modern ultrafast spectroscopic tools became available that the search for nontrivial quantum effects in the primary steps of biological processes was made possible, which led to the prospect for the foundation of the field of quantum biology \cite{ Nat_Phys_9_10_(2013), Nature_543_647_2017, Science_340_1448_2013, Nature_543_355_2017}. Recent experimental results obtained for the well-characterized Fenna-Mathews-Olson (FMO) protein complex have been interpreted as evidence of long-lived electronic quantum coherence in the primary steps of the energy transfer \cite{Nature_446_782_2007, Nature_463_644_2010, Nat_Chem_4_389_2012, Chem_Phys_386_1_2011}. A functional role of long-lived quantum coherence was proposed in that it would speed up the transfer of excitation energy under ambient conditions \cite{PNAS_106_17255_2009}. These works triggered tremendous interest in different fields, ranging from quantum chemistry to  quantum information science. A key parameter is the strength of the coupling of the exciton system to environmental fluctuations, which is related to the reorganization energy. For a conceptual understanding, initial theoretical analysis was built upon the choice of a rather small reorganization energy of 35 cm$^{-1}$ to fit the reported long lifetime of electronic coherence \cite{PNAS_106_17255_2009}. Yet, even with this small value, Shi and coworkers found a shorter lifetime for the expected electronic coherence from more advanced calculations of the experimental 2D electronic spectra  \cite{JCP_134_194508_2011}. The interpretation of the long lifetime of the electronic coherence was further questioned by numerically exact results obtained within the quasi adiabatic propagator path integral method with an experimentally determined spectral density with a considerably larger reorganization energy \cite{PRE_84_041926_2011}. Coker {\em et al.} and Kleinekath{\"o}fer {\em et al.} have calculated site-dependent reorganization energies with refined atomic details by  advanced molecular dynamics simulations \cite{JPCL_7_3171_2016, JPCB_119_9995_2015}. They found significantly larger values of the reorganization energies in the range of 150 to 200 cm$^{-1}$. With this disagreement, we have revisited the energy transfer of the FMO complex at room temperature experimentally \cite{PNAS_114_8493_2017} using 2D electronic spectroscopy to extract the electronic coherence time scales. Instead of a long-lived electronic coherence, the experiment, after having passed a self-consistency verification, yielded a considerably shorter coherence lifetime of 60 fs. This observed timescale for decoherence excludes any functional role for coherent energy transfer in the FMO complex, which occurs on the time scale of several picoseconds at room temperature. 

Another potential key role for the electronic coherence is played by the pigment-protein host molecular vibrations  \cite{JPCA_118_10259_2014, CPL_545_40_2012, JPCC_117_18728_2013, JPCL_3_2828_(2012)}. In contrast to electronic coherence, the pigment-localized vibrations typically last for picoseconds but are not expected to enhance energy transfer in general. Yet, Plenio and coworkers have suggested the concept of vibrationally enhanced electronic coherence \cite{Nat_Phys_9_113_2013}. They reported that in a vibronic model dimer, electronic quantum coherence may be resonantly enhanced by long-lasting vibrational coherence \cite{JCP_139_235102_2013}. Instead of an enhancement, Tiwari {\em et al.} alternatively suggested that nonadiabatic electronic-vibrational mixing may resonantly enhance the amplitude of particular, delocalized anticorrelated vibrational modes of the electronic ground state \cite{PNAS_110_1203_(2013)}. While in principle, this mechanism is also possible in the presence of weak electronic dephasing \cite{Yeh}, realistic values of the strengths of the electronic and vibrational dampings leads to a complete suppression of this mechanism \cite{DuanJCP}. Electronic-vibrational mixing was also examined in a simple dimer  model \cite{Nat_Chem_6_196_2014}, but the subsequent theoretical calculations show no evidence of an enhancement of the electronic coherence \cite{NJP_17_072002_2015}. More recently, the coherent exciton transfer in the FMO complex has been revisited by Zigmantas and coworkers at 77 K \cite{Nat_Chem_10_780_2018}. The long-lived oscillations have been carefully assigned to the vibrational coherence of the electronic ground state. Due to strong dissipation, the lifetime of the electronic coherence was too short to be precisely determined even at 77 K. Despite the extensive work on this problem, a complete picture of the electronic coherence and its role in the electronic-vibrational mixing for the energy transfer in the FMO complex is still elusive. 

Here, we study the energy transfer process in the FMO complex with the explicit aim to observe clear evidence for the onset of electronic coherence effects in energy transport by decreasing temperature. For this, we measure the 2D electronic spectra of the FMO complex in the regime of very low temperature. Specifically, we examine the electronic dephasing by directly measuring the anti-diagonal bandwidth of the main peaks, along with the decays in the cross peaks related to the inter-exciton coupling.  It is only at very low temperatures (20 K) that the amplitudes and decays in these electronic coherence signatures become comparable to the energy transfer times. We provide a comprehensive analysis by a global fitting approach, and the subsequent Tukey window Fourier transform allows us to disentangle the electronic coherence from vibrational coherence. By this, we uncover that the longest lived electronic coherence is observable up to 500 fs between the two excitons closest to the reaction centre side. Due to down-hill energy transfer, the electronic coherence of the two higher-energy excitons close to the antenna side shows a much faster decay with a lifetime $<$60 fs. We furthermore measure the coherent energy  transfer over an extensive temperature range. Based on these temperature-dependent measurements, we are able to construct a unifying exciton model, which captures the coherent energy transfer over the entire temperature range studied. Moreover, we investigate the temperature-dependent non-Markovianity of the transfer dynamics to show that the bath fluctuations are uncorrelated even at low temperatures. By this unprecedented combination of experimental and theoretical efforts, we are able to provide a complete picture of quantum coherent effects in the FMO complex over the entire regime from high to low temperatures in one experiment and one theoretical model. Due to the generic structure of the FMO protein, we expect our observations to be extended to other more complicated photosynthetic protein complexes and even photovoltaic devices \cite{Kirk2018}.

%%%%%%%%%%%%%%%%%%%   Results and Discussion 
\section*{Results} 

The solution of the FMO protein complex is prepared in a home-built sample cell and mounted in a cryostat (Oxford Instrument). More details of the sample preparation are given in the Materials and Methods section. Fig.\ \ref{fig:Fig1}(a) shows the structural arrangement of the bacteriochlorophyll a (Bchla) chromophores embedded in the protein matrix (data from 3ENI.pdb). The measured absorption spectrum of the FMO complex at 80 K and the laser spectrum used in this study are shown in the SI. 

\subsection{Two-dimensional electronic spectroscopy} 

\begin{figure}[h!]
\begin{center}
\includegraphics[width=18.0cm]{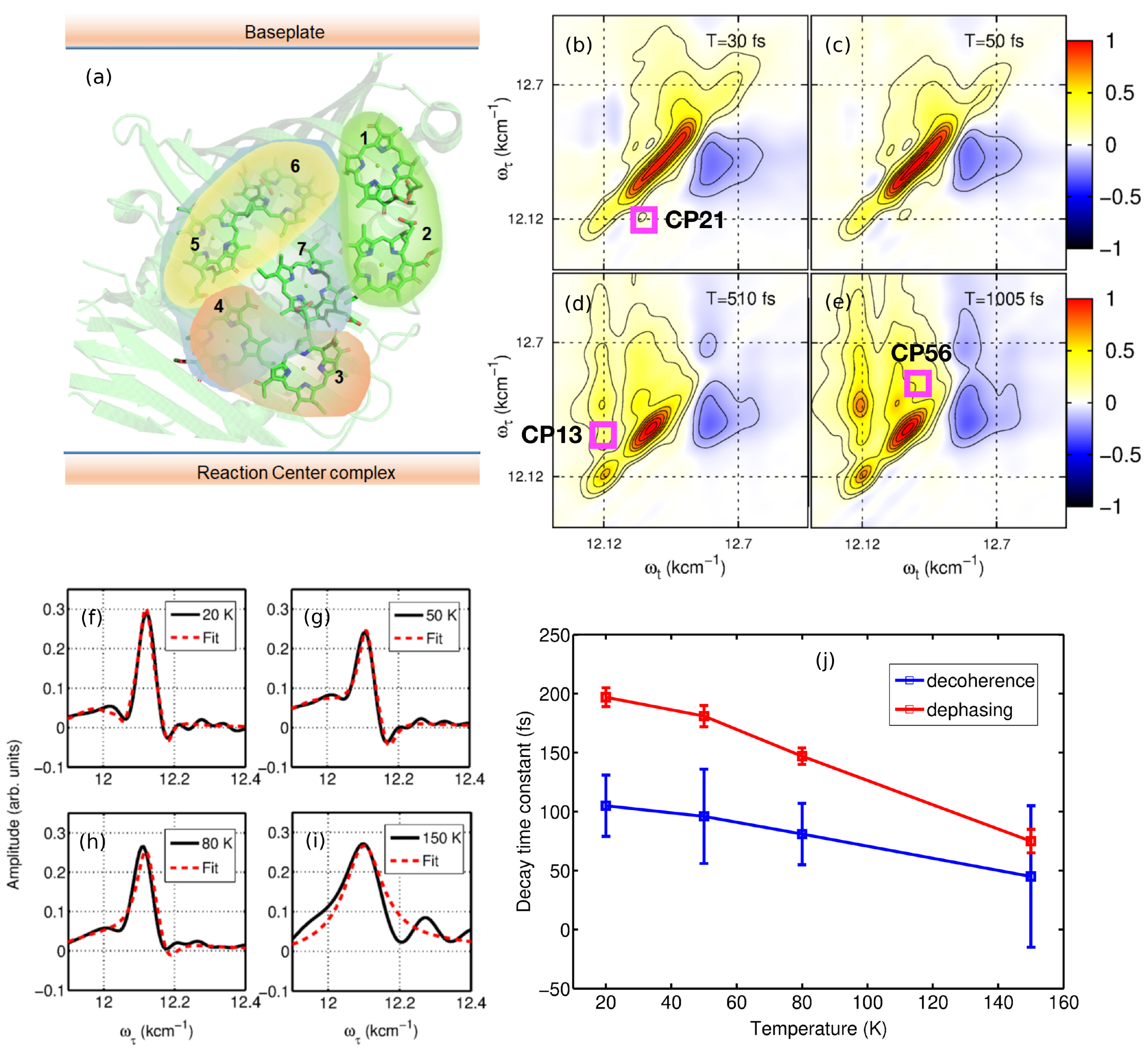}
\caption{\label{fig:Fig1} (a) The structural arrangement of the pigments in the FMO protein complex. (b-e) Real parts of measured 2D electronic spectra of the FMO complex  for the waiting times 30, 50, 510 and 1005 fs, respectively. (f-i) Anti-diagonal bandwidth of the diagonal peak at ($\omega_{\tau}$, $\omega_{t}$) = (12120, 12120) cm$^{-1}$ at 20, 50, 80 and 150 K. The decay times of the electronic dephasing between the ground and excited states are obtained by fitting to a Lorentzian lineshape (red dashed line). We find 197, 181, 147 and 75 fs, respectively. These values (``dephasing'') are shown in (j), together with the decay times associated with the cross-peak decay for growing waiting times.} 
\end{center}
\end{figure}

We measure the 2D electronic spectra of the FMO complex in the range from 20 to 150 K. The details of the 2D spectrometer are given in the Materials and Methods section. The real parts of the 2D electronic spectra at 20 K for selected waiting times T are shown in Figs.\ \ref{fig:Fig1}(b-e). The positive and negative amplitudes in the peaks represent the excitation transitions of the ground-state-bleach (GSB) and the excited-state-absorption (ESA), respectively. In Fig.\ \ref{fig:Fig1}(b), we show the measured 2D electronic spectrum at 20 K for T=30 fs. The exciton states in the FMO complex are located in the frequency range from 12120 to 12700 cm$^{-1}$, which is marked by black dashed lines. At T=30 fs, we observe a dramatic stretch of the main peaks along the diagonal, which illustrates the strong inhomogeneous broadening. In addition, one off-diagonal feature corresponding to the ESA is observed at ($\omega_{\tau}$, $\omega_{t}$) = (12300, 12580) cm$^{-1}$. At T=50 fs as shown in Fig.\ \ref{fig:Fig1}(c), we observe that the 2D spectrum has not changed significantly, except that the elongation of the main diagonal peaks is slightly reduced. However, the elongation of the main peaks along the diagonal has dramatically reduced again at T=510 fs in Fig.\ \ref{fig:Fig1}(d). Moreover, we observe that the main peaks of the higher exciton states are replaced by one peak with ESA features. A new cross peak appears at ($\omega_{\tau}$, $\omega_{t}$) = (12340, 12120) cm$^{-1}$, which provides evidence of the down-hill energy transfer from higher exciton states to the lowest ones. Its amplitude  is further increased at T=1005 fs, see Fig.\ \ref{fig:Fig1}(e). In addition to the main and cross peaks in the frequency range from 12120 to 12700 cm$^{-1}$, we observe more cross peaks appearing at the upper-left side of the 2D electronic spectrum at T=1005 fs.  This provides evidence of the vibrational progression in the FMO complex. To examine the lifetime of the electronic dephasing, we analyze the anti-diagonal bandwidth of the lowest exciton peak at ($\omega_{\tau}$, $\omega_{t}$) = (12120, 12120) cm$^{-1}$ for T=30 fs. To quantify the associated lifetime of the optical coherence, the broadening of the peaks is modeled by Lorentizan lineshapes. More details of the fitting procedure are described in the SI. By this, we are able to capture the electronic dephasing between the electronic ground and excited states for different temperatures, see Figs.\ \ref{fig:Fig1}(f)-(i). We find the lifetimes of the electronic dephasing of 197, 181, 147 and 75 fs for the temperatures 20, 50, 80 and 150 K, respectively, which are shown in Fig.\ \ref{fig:Fig1}(j), marked by ``dephasing''.

\subsection{Energy transfer and coherent dynamics} 

\begin{figure}[h!]
\begin{center}
\includegraphics[width=15.0cm]{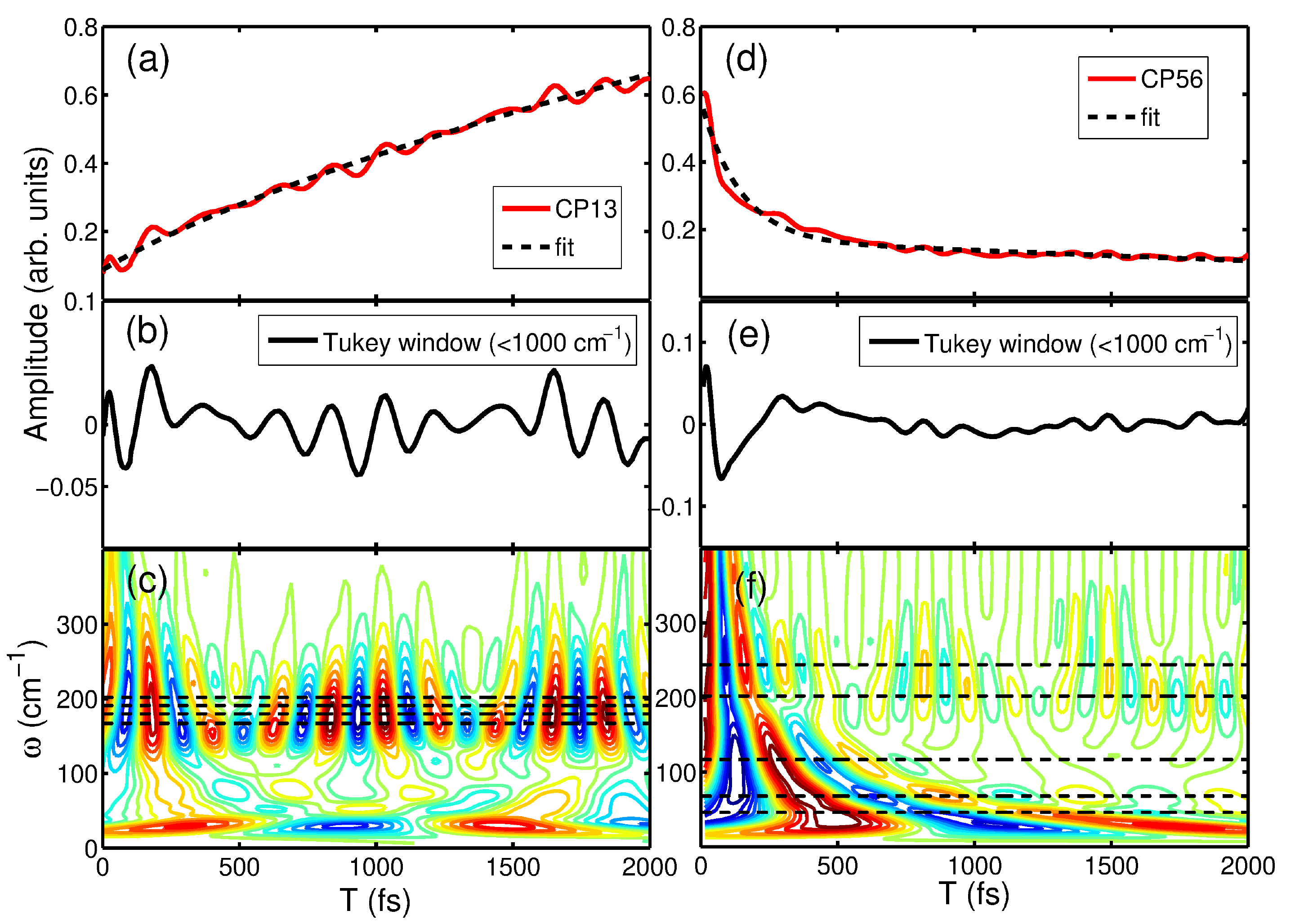}
\caption{\label{fig:Fig2} (a) Kinetics (red line) and fitting trace (black dashed line) of the cross peak measured at ($\omega_{\tau}$, $\omega_{t}$) = (12340, 12120) cm$^{-1}$. (b) The time trace (black solid line) after removing high-frequency noise by a Tukey window Fourier analysis. (c) Time evolution of the vibrational coherences in a wavelet analysis. Close-by lying frequencies of the ground-state vibrational modes at 167, 180, 191 and 202 cm$^{-1}$ result in an oscillatory resonant beating at the waiting times T = 1000 and 1700 fs. (d) Kinetics (red line) and fitting curve (black dashed line) of the cross peak at ($\omega_{\tau}$, $\omega_{t}$) = (12570, 12480) cm$^{-1}$. (e) The time trace after Tukey-window treatment (black solid line). (f) Wavelet spectrum with the coherent dynamics of the vibrational modes at 46, 68, 117, 202 and 243 cm$^{-1}$.  }
\end{center}
\end{figure}

To examine the time-dependent coherent dynamics in the 2D spectra, we extract the magnitudes of the cross peaks at different waiting times. In Fig.\ \ref{fig:Fig2}(a), the trace (red line) represents the time evolution of the amplitude of the cross peak at  ($\omega_{\tau}$, $\omega_{t}$) = (12340, 12120) cm$^{-1}$  between exciton 1 and 3 (marked as `CP13' in Fig.\ \ref{fig:Fig1}(d)). The underlying kinetics (black dashed line) is fitted by an exponential function and the resulting residual is shown as a black solid line in Fig.\ \ref{fig:Fig2}(b). The raw data of the oscillations are further purified by a Fourier filter with a Tukey window ($<$1000 cm$^{-1}$) in Fig.\ \ref{fig:Fig2}(b). With this refined trace, we retrieve the coherent dynamics by performing a wavelet analysis. The details of the Tukey window Fourier transform and the wavelet analysis are given in the SI. The time evolution of the cross-peak coherence is shown in Fig.\ \ref{fig:Fig2}(c). We are able to resolve oscillations with frequencies in the range from 160 to 200 cm$^{-1}$. Clearly, these identified mode frequencies are in excellent agreement with the vibrational modes in the FMO complex identified by fluorescence-line-narrowing (FLN) studies \cite{JL_127_251_2007}. More specifically, we recover the modes at 167, 180, 191 and 202 cm$^{-1}$ on the electronic ground state, which are highlighted by black dashed lines in Fig.\ \ref{fig:Fig2}(c). Since the frequencies lie close-by, we observe clear evidence of beating of vibrational oscillations at waiting times between 1000 and 1700 fs. Our analysis clearly reveals that the beatings  originate from the resonant enhancement of individual vibrational (electronic ground state) modes at close-by frequencies. These vibrational beatings have been incorrectly assigned to the enhancement of electronic coherence by long-lasting vibrational coherences. However, our analysis now uncovers that this enhancement is purely caused by resonant beating of vibrational molecular modes in the electronic ground state. Moreover, we show the traces of the cross peak amplitudes (red solid line) at ($\omega_{\tau}$, $\omega_{t}$) = (12570, 12480) cm$^{-1}$, CP56 (marked in Fig.\ \ref{fig:Fig1}(e)), the associated fits (black dashed line) in Fig.\ \ref{fig:Fig2}(d)., and the polished residuals (black solid line) stemming from the Tukey window Fourier transforms in Fig.\ \ref{fig:Fig2}(e). We plot the obtained results of the wavelet analysis, which yield the coherent dynamics of the residuals in Fig.\ \ref{fig:Fig2}(f). The resolved modes are marked as black dashed lines. We find the mode frequencies of 46, 68, 117, 202, 243 cm$^{-1}$, which perfectly agree with those revealed by FLN experiments \cite{JL_127_251_2007}. 

Having analyzed the vibrational coherences, we next examine the time scales and the pathways of the energy transfer by the global fitting approach \cite{global_fitting}. First, we construct a 3D data set by combining a series of 2D electronic spectra with evolving waiting times. At least two exponential functions have been used to achieve a converged fitting. This yields the time constants of the energy transfer  of 160 fs and 8.8 ps, respectively. The corresponding two decay-associated spectra are shown in Figs.\ S6 and S7 of the SI. It is interesting that the fastest component of the energy transfer is 160 fs and thus close to the retrieved life time of the  electronic coherence (described in the next section). Hence, at this very low temperature, this dynamical component of the energy transfer may largely be mediated by electronic quantum coherence. The ultrafast timescales and the multiple pathways of the energy transfer are described in detail in the SI. Moreover, we also show in the SI the other decay-associated spectrum related to the time constant of 8.8 ps. In particular, this shows clear evidence of a down-hill energy transfer from higher-energy excitons to the lowest-energy exciton.

\subsection{Electronic quantum coherence} 

\begin{figure}
\begin{center} 
\includegraphics[width=15.0cm]{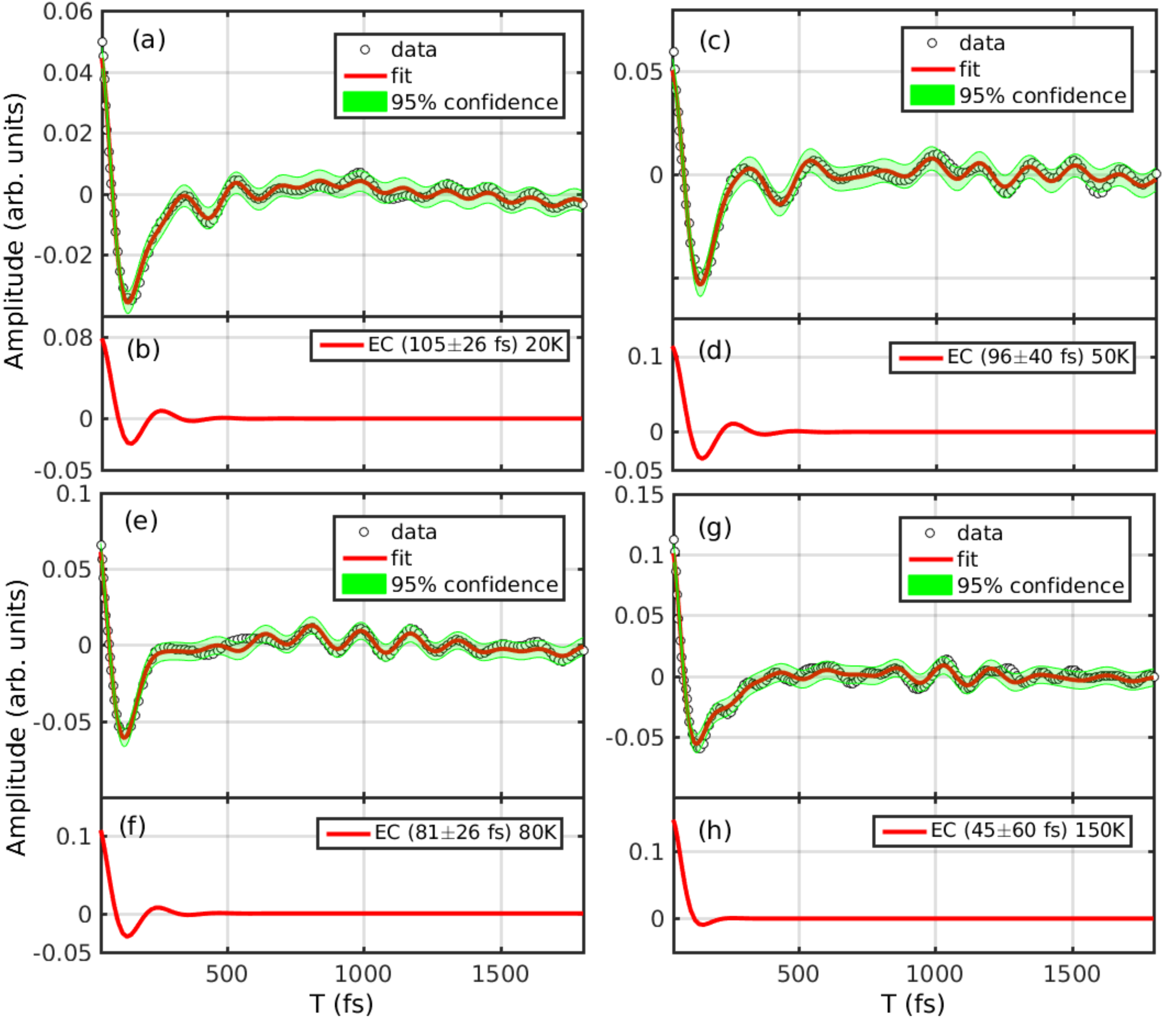} 
\caption{\label{fig:Fig3} Lifetime of the measured electronic quantum coherence between exciton 1 and 2 at different temperatures. (a) Residual (black circles) of the cross peak (20 K) at ($\omega_{\tau}$, $\omega_{t}$) = (12120, 12270) cm$^{-1}$ and fit to a multi-exponential function (red line). The green shadow shows the boundaries of 95$\%$ confidence. (b) Resulting electronic coherence  with a decay time constant of 105$\pm$26 fs. The oscillation frequency of 150 cm$^{-1}$ perfectly matches the energy gap between exciton 1 and 2. (c), (e) and (g) Residuals and fitting traces at 50, 80 and 150 K, obtained following the same procedure. (d), (f) and (h) The obtained electronic coherences with decay time constants of 96$\pm$40, 81$\pm$26 and 45$\pm$60 fs. }
\end{center}
\end{figure}

To capture the signature of electronic coherence, we address the cross-peak dynamics associated to the two excitons 1 and 2 with lowest energy. We take the time-dependent magnitude of the cross peak at ($\omega_{\tau}$, $\omega_{t}$) = (12120, 12270) cm$^{-1}$ (marked as `CP21' in Fig.\ \ref{fig:Fig1}(b)) to minimize contributions by the energy transfer dynamics. Again, the residual obtained after removing the kinetics  and after polishing by a Tukey window Fourier transform is shown as black circles in Fig.\ \ref{fig:Fig3}(a). Here, the oscillations are induced by the electronic coherence convoluted with vibrational coherences. To disentangle both, the oscillations and decay rates are fitted to exponentially decaying sine functions to extract the oscillation frequencies and the lifetimes of the coherences. We start the fit by assuming the frequencies 68, 150, 180 and 202 cm$^{-1}$, which represent the vibrational modes found experimentally as discussed above and which agree with the known modes from the FLN experiment \cite{JL_127_251_2007}. Moreover, we have obtained the electronic energy gap of 150 cm$^{-1}$ between exciton 1 and 2 by theoretical calculations (see below), which also agrees with previous results \cite{JPCL_7_1653_2016}. In addition, to remove the low-frequency oscillations, one additional frequency of 17 cm$^{-1}$ is included to achieve the best fit with R-square $>$ 0.97. It represents the lowest frequency resolved associated to our basic time step. All the fitting procedures are performed using the Curve Fitting Toolbox in Matlab 2013(b); the details are given in the SI. We show the high-quality fitting results by the red solid line in Fig.\ \ref{fig:Fig3}(a). The green shadow indicates the boundaries of 95$\%$ of fit confidence. This effectively allows us to separate the electronic coherence of 150 cm$^{-1}$ from vibrational coherences. The oscillation related to the electronic coherence is shown in Fig.\ \ref{fig:Fig3}(b) and yields a decay time constant of 105$\pm$26 fs. With the frequency of 150 cm$^{-1}$, it is clearly observed that the electronic coherence is sustained over only two oscillation periods and disappears within 500 fs completely. More important, we observe that the identified oscillations of the electronic coherence are quite significant. They are larger than 5$\%$ of the maximum strength of the 2D spectra at 20 K. In addition, the identified vibrational coherences are shown in the SI. Following the same procedures, we analyze the coherent dynamics of the cross peak at ($\omega_{\tau}$, $\omega_{t}$) = (12120, 12270) cm$^{-1}$ (CP21) at different temperatures (50, 80 and 150 K) and plot the corresponding traces in Fig.\ \ref{fig:Fig3}(c), (e) and (g), respectively. The decay time constants for different temperatures are shown in Fig.\ \ref{fig:Fig1} (j) marked as "decoherence". The resolved extracted electronic coherences are shown in Fig.\ \ref{fig:Fig3}(d), (f) and (h), respectively. Noticeably, at 50 K, the electronic coherence lasts less than 500 fs, with a decay time constant 96 $\pm$ 40 fs. The lifetime of the electronic coherence is significantly reduced at 80 K. We measure a decay time constant of 81 $\pm$ 26 fs. At 150 K, the red solid line in Fig.\ \ref{fig:Fig3}(h) clearly shows that the electronic coherence is strongly damped and the oscillation does not survive even a single oscillation period. Again, the associated vibrational coherences retrieved during the fitting procedures are shown in the SI. 

\begin{figure}
\begin{center} 
\includegraphics[width=17.0cm]{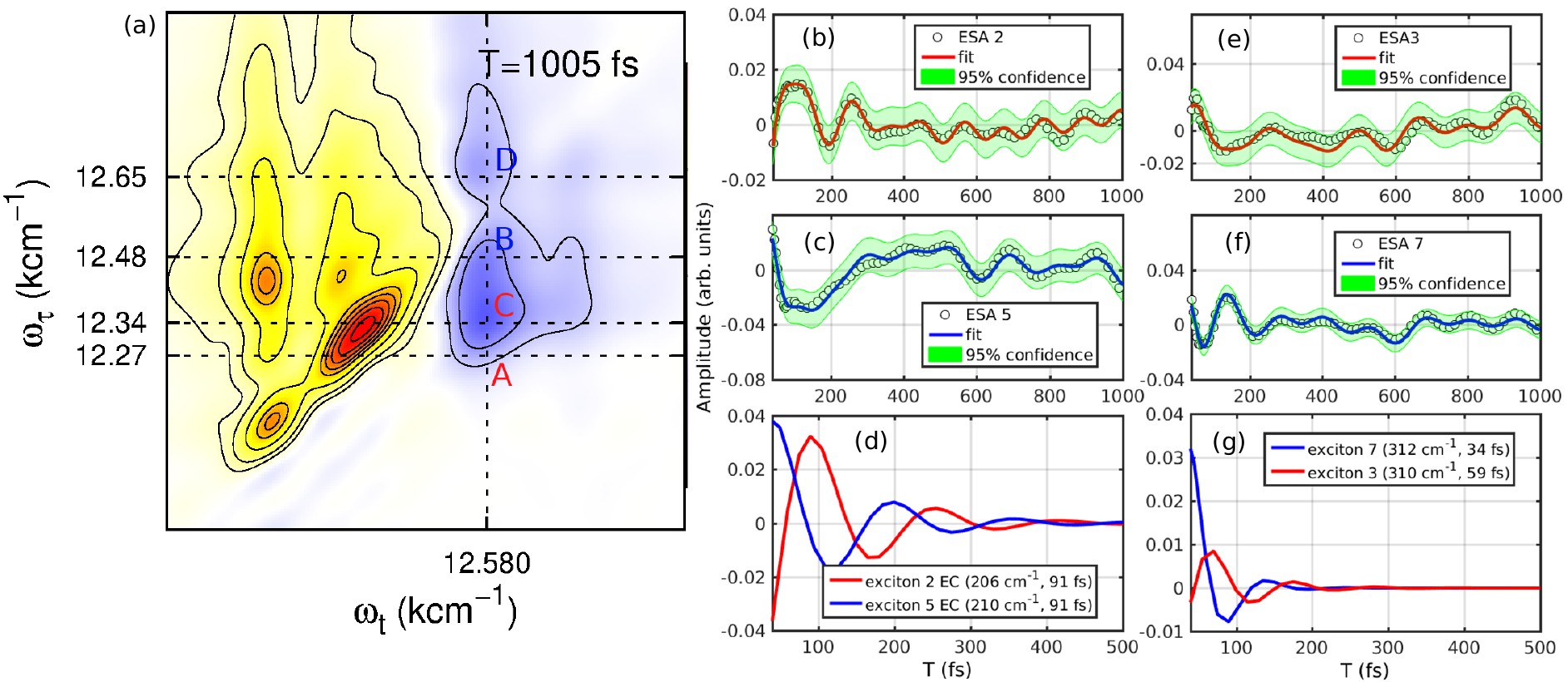} 
\caption{\label{fig:Fig4} (a) Measured 2D electronic spectrum at T=1005 fs. The ESA peaks of exciton 2, 3, 5 and 7 are labeled as A, C, B and D, respectively. (b) Residual of the ESA peak (black circles) of exciton 2 (red marker A, $\omega_{\tau}=12270$ cm$^{-1}$) after removing the kinetics by exponential fits. The residual is further analyzed by fitting functions (red line, for details of the fitting procedure, see the Supporting Information). Following the same procedure, the residual (black circles) and the fitting trace (blue line) for exciton 5 are shown in (c) (blue marker B in (a), $\omega_{\tau}=12480$ cm$^{-1}$). (d) The obtained oscillations of the electronic coherence between exciton 2 and 5 with the frequencies of 206 and 210 cm$^{-1}$. The decay constants are 91 fs. The frequencies agree perfectly with the energy gap between exciton 2 and 5. (e) and (f) Residuals and fitting traces of exciton 3 and 7. (g) The obtained electronic coherences of exciton 3 and 7 with the decay time constants of 34 and 59 fs, respectively. The resolved oscillations show the frequencies of 312 and 310 cm$^{-1}$, which matches the energy difference between exciton 3 and 7. } 
\end{center} 
\end{figure}

To study the coherence of excitons in states of higher energy, we monitor the dynamics of ESA peaks at $\omega_{t}$ = 12580 cm$^{-1}$. Our theoretical calculations (see details in the next section) retrieved energy levels of the excitons 2, 3, 5 and 7 highlighted by the corresponding $\omega_{\tau}$-lines in Fig.\ \ref{fig:Fig4}(a). The intersection points of two marker lines are denoted as A, B, C and D, respectively, to which the excited state dynamics of the excitons 2, 3, 5 and 7 corresponds. Following the same procedure as above, we extract the time evolution of the amplitude of the cross peaks at A, B, C and D and remove the underlying kinetics by the global fitting approach. The obtained residuals of these peaks after the Fourier treatment are shown in Fig.\ \ref{fig:Fig4}(b), (c), (e) and (f), respectively, for increasing waiting times. %To achieve a good fit, we repeat the fitting procedure with a function of sine and exponential functions. The initial guessed frequencies of electronic and vibrational coherences are based on the identified peaks after Fourier transform. 
More details of the fitting procedure are presented in the SI. By this, a high-quality fit is obtained, which is shown as the red solid line in Fig.\ \ref{fig:Fig4}(b). The extracted electronic coherence is presented as red solid line in Fig.\ \ref{fig:Fig4}(d). We find the frequency of 210 cm$^{-1}$ of this coherent oscillation, which is in excellent agreement with the energy gap between exciton 2 and 5 from our theoretical calculations. Next, we examine the subsequent coherent dynamics of the ESA peak B. From our theoretical analysis, we expect this to be a peak that originates from the ESA of exciton 5. After repeating the fitting procedure described above, we obtain the residual shown in Fig.\ \ref{fig:Fig4}(c). The measured residuals and the results of the fits are shown as black dots and blue solid line, respectively. Furthermore, we show the separate electronic coherence as blue solid line in Fig.\ \ref{fig:Fig4}(d). Again, the obtained frequency of 206 cm$^{-1}$ matches the energy gap between exciton 5 and 2 exactly. Interestingly, these well-resolved electronic coherences in Fig.\ \ref{fig:Fig4}(d) show evidence of anti-correlated oscillations with a slight phase offset. From our theoretical analysis, we uncover the coherence between exciton 2 and 5 is dominated by the strong electronic couplings of pigment 4 and (5, 6). Details of the transformation from the site to the exciton basis are given in the SI. The residuals of the ESA peaks C and D are shown as black dots in Fig.\ \ref{fig:Fig4}(e) and (f), respectively. The fits to the oscillations are shown as red and blue solid lines in Fig.\ \ref{fig:Fig4}(e) and (f) and the extracted electronic coherences are shown as red and blue solid lines in Fig.\ \ref{fig:Fig4}(g). We observe two electronic coherent oscillations both with the frequency of $\sim$310 cm$^{-1}$, which agrees perfectly with the energy gap between exciton 3 and 7. As determined by the basis transformation, this coherence is dominated by the electronic coupling between pigment 1 and 2. Moreover, compared to the lifetime of the coherence between  exciton 2 and 5, these coherences show  smaller time constants of 34 and 59 fs. Hence, the electronic coherence lasts shorter for higher exciton states, which is due to the down-hill energy transfer. Moreover, the larger energy gap between exciton 3 and 7 produces a shorter lifetime of the electronic coherence due to the faster energy transfer.

\subsection{Theoretical calculations} 

\begin{figure}
\begin{center} 
\includegraphics[width=15.0cm]{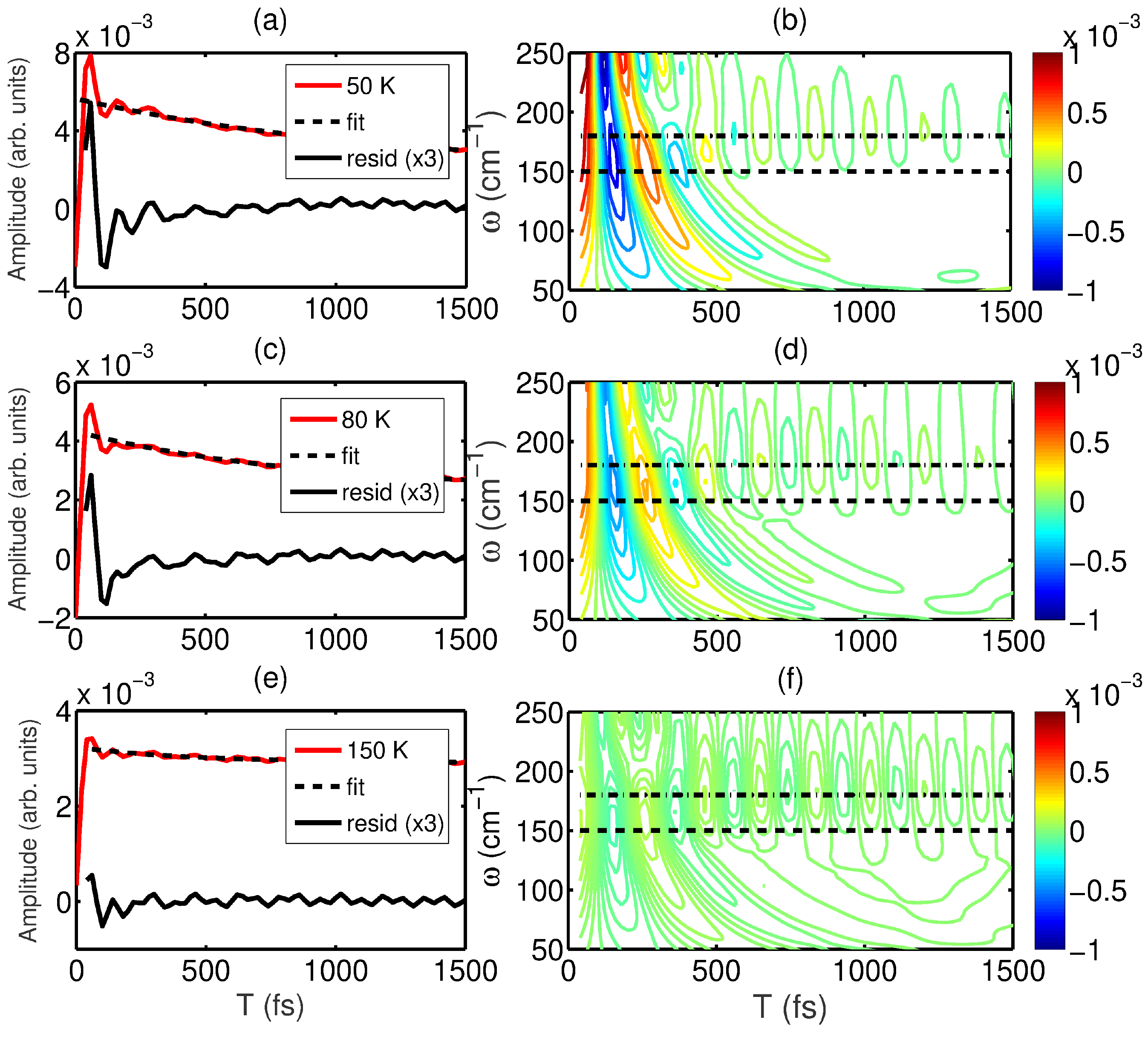} 
\caption{\label{fig:Fig5} (a) Time trace (red solid line) of the cross peak between exciton 1 and 2 (CP21) in the calculated 2D electronic spectra. The residuals (magnified by 3 times) are plotted as black solid line after removing the kinetics (black dashed line) by an exponential fit. (b) Wavelet analysis of the residuals at 50 K. The frequencies of the coherent dynamics of 150 and 180 cm$^{-1}$ are highlighted by black dashed and black dashed-dotted lines, respectively. (c) Trace (red solid line) of the cross peak between exciton 1 and 2 at 80K. The subsequent fit and the residuals are shown as black dashed and solid lines. The frequencies of the coherent dynamics of 150 and 180 cm$^{-1}$ are marked in (d). (e) Trace of the cross peak between exciton 1 and 2 at 150 K. The fit and residuals are shown as black dashed and solid lines, the wavelet spectrum is shown in (f). } 
\end{center}
\end{figure}

We construct a Frenkel-exciton model to study the coherent FMO dynamics. The electronic transitions in the pigments are approximated by optical transitions between two energy eigenstates and the electronic couplings between pigments are calculated within the dipole approximation. Moreover, to include fluctuations of the electrostatic interactions, the system is linearly coupled to a  harmonic reservoir. For simplicity, we consider a standard Drude bath spectral density. Moreover, a particular localized vibrational mode of 180 cm$^{-1}$ is coupled to the electronic system to investigate the role of vibrational/vibronic coherence. This vibrational mode is known to be the most relevant out of a group of 44 vibrational modes. The 2D electronic spectra are calculated by a time non-local quantum master equation \cite{TNL1, TNL2} and the time evolution of the peaks are obtained by the equation-of-motion phase-matching approach \cite{EOMPMA}. More details are given in the Materials and Methods section and the SI. We initially choose the site energies from previous works \cite{JPCL_7_1653_2016} and then optimize them by simultaneously fitting to the experimental absorption spectra  measured at different temperatures. After that, we calculate the 2D electronic spectra and refine then the system-bath interaction strength by comparing the calculated electronic dephasing lifetimes to the experimental ones at different temperatures. By this, we are able to develop a unique system-bath model with a single set of parameters. In particular, we are now able to calculate the 2D spectra. 
We show the calculated waiting time traces of the cross peak at ($\omega_{\tau}$, $\omega_{t}$) = (12120, 12270) cm$^{-1}$ (CP21) in Fig.\ \ref{fig:Fig5} which reveal the coherent dynamics between exciton 1 and 2. The obtained red solid lines at 50, 80 and 150 K are plotted in Fig.\ \ref{fig:Fig5}(a), (c) and (e), respectively. As above,  we analyze the calculated data again by the global fitting approach. The so obtained fits of the traces are plotted as black dashed lines and the retrieved residuals are shown as black solid lines in Fig.\ \ref{fig:Fig5}(a), (c) and (e). To calculate the time-dependent cross-peak dynamics, we use the wavelet analysis aiming to separate the electronic and vibrational coherences.  The results are shown in Fig.\ \ref{fig:Fig5}(b), (d) and (f). We find, at 50 K, clear evidence of long-lived vibrational coherence at 180 cm$^{-1}$, as marked by the dashed-dotted line in (b). In addition to the vibrational mode frequency,  clear evidence of electronic coherence is highlighted in the wavelet spectrum by the black dashed line in (b). We observe that the electronic coherence lasts for two oscillation periods and has disappeared at 500 fs. More importantly, based on the wavelet analysis, the oscillation phase retrieved in (b) perfectly matches that revealed by the experimental data shown in Fig.\ \ref{fig:Fig3}(d). More details of the comparison are shown in the SI. Moreover, the electronic (black dashed line) and vibrational (black dash-dot line) coherences at 80 and 150 K are shown in Fig.\ \ref{fig:Fig5}(d) and (f). On the basis of the results at different temperatures, we can safely conclude that the lifetime of the vibrational coherence at 180 cm$^{-1}$ is not significantly shortened for increasing temperature. However, the lifetime of the electronic coherence is dramatically reduced when temperature is increased. The same conclusion can be drawn for the higher-energy excitons at different temperatures; the details of the coherent lifetimes of higher excitons are shown in the SI. 

\section*{Discussion} 

An important question related to the exciton transfer dynamics is about the nature of the bath-induced fluctuations. It may be characterized by the measure of the non-Markovianity which quantifies how strongly the dephasing and relaxation dynamics departs from ordinary Markovian, i.e., memory-less behaviour. In principle, a highly structured environment consisting of several localized vibrational modes of the FMO protein may give rise to significant non-Markovian dynamics. Early numerically exact path-integral calculations \cite{PRE_88_062719_2013} on the basis of an experimentally determined spectral density have shown that the exciton dynamics is purely Markovian at ambient conditions. This expectation has been confirmed recently experimentally by comparing the decay time of optical dephasing and electronic quantum coherence in the  FMO complex \cite{PNAS_114_8493_2017}. The equivalence of the time scales of the optical dephasing and the electronic decoherence reveals that the dynamics of energy transfer in the FMO complex at room temperature is fully Markovian. 

Here, we provide a complete picture of the role of the non-Markovianity in the FMO complex at different temperatures. As discussed above, we have measured, at 20 K, the longest life time of the electronic quantum coherence for the dephasing of the two lowest-energy excitons, with the decay time of 105 fs. On the other hand, the analysis of the anti-diagonal band width yielded a decay time of 197 fs for the optical dephasing of excitons 1. The difference of almost a factor of 2 is due to the low temperature, but is still covered by a fully Markovian description of the transfer dynamics \cite{Sorgenfrei}. This finding is also in agreement with low-temperature calculations \cite{PRE_88_062719_2013}. On the basis of these studies, we conclude that non-Markovian energy fluctuations of the pigments induced by pigment-hosted molecular vibrations do not play a role in the energy transfer of the FMO complex. Even, due to its simple structure, we believe that this conclusion can be extended to more complicated photosynthetic protein complexes.  

Instead of long-lived electronic coherence, our study uncovers a long lasting beating dynamics composed of vibrational coherences in the range of 180 cm$^{-1}$ in the electronic ground state. As summarized in the Introduction, several studies have suggested a resonant enhancement of the short-lived electronic coherence by the long-lived vibrational coherence in the FMO complex. However, our work clearly shows no such a resonant enhancement of the electronic coherence during the population transfer.  In contrast, from the measured and calculated 2D electronic spectra, we have retrieved the scale of the reorganization energy of 120 cm$^{-1}$, which manifests a quite  strong system-bath interaction that rapidly destroys the phase of the electronic quantum coherences between pigments in the FMO complex. Hence,  instead of the energy transfer being enhanced by strongly delocalized exciton wave functions, the rather large reorganization energy sharpens the limits for delocalization and the energy pathways between pigments. Then, the efficient down-hill transfers of not too strongly delocalized excitons are determined by a simple thermal distribution of excitons which rapidly arises after an initial ultrafast non-equilibrium dynamics triggered by the photo-excitation. 

\section*{Conclusions} 

In this paper, we provide a complete picture of the coherent contribution to energy transfer in the FMO complex by 2D electronic spectroscopy in the entire regime from low to high temperatures. In particular, the spectroscopic measurements at low temperature of 20 K allows us to provide unambiguous evidence of the lifetime of the electronic quantum coherence and to disentangle the electronic coherence from long-lived vibrational coherence. Interestingly, due to the down-hill energy transfer, the electronic coherence between the two lowest excitons is marginally observable out to 500 fs at 20 K. However, the coherence lifetime of higher excitons is dramatically reduced by the population transfers. This analysis allows us to disentangle the previously reported long-lived beating of cross-peak signals and to show that they are composed of mixed ground state molecular Raman modes.  Moreover, we uncover that the lifetime of electronic coherence is significantly modulated by temperature, while, in contrast, the resonant beatings of vibrational coherences last for picoseconds even at 150 K. A thorough analysis on the basis of a unique combination of experimental data and theoretical modelling enables us to provide a reliable estimate for the decisive parameter of the reorganization energy of the FMO complex. We find a reorganization energy of 120 cm$^{-1}$, which represents a strong system-bath interaction of the pigments with their protein environment. This coupling is sufficient to significantly reduce the lifetime of electronic coherences and leads to a rapid intermittent localization of the electronic wavefunction on a few molecular sites. Instead of a long-lived quantum coherent energy transfer, we provide a different picture of a down-hill energy transfer, in which the pathways of population transfers are dynamically constructed simply by following lower site energies of the pigments involved and by rather fast electronic damping due to significant nuclear reorganization in the excited state. The latter sharpens the funnels of the energy flow in the FMO complex. In general, we conclude that the energy transfer in the FMO complex is dominated by the thermal dynamics of weakly delocalized excitons after an initial ultrafast non-equilibrium photon-excitation. Due to common features in the bath for all light harvesting systems, despite the relative simplicity of FMO, we believe that this conclusion can be further extended to the other more complicated photosynthetic protein complexes. 

%\end{Conclusions}
%

%%%%%%%%%%%%%%%%%%     M E T H O D S

\section*{Materials and Methods}
\subsection{Sample preparation.}

The FMO protein was isolated from the green sulfur bacteria {\em C. tepidum} (see the SI for more details). The sample was dissolved in a Tris buffer at PH 8.0. It was filtered with a 0.2 $\mu$m filter to reduce light scattering. The sample was then mixed 70:30 v/v in glycerol and kept in a home-built cell with an optical pass length of 500 $\mu$m. The cell was mounted in the cryostat (MicrostatHe-R) for the low-temperature measurements. 

\subsection{2D Electronic measurements with experimental conditions.} 

Details of the experimental setup have already been described in earlier reports from our group \cite{PNAS_114_8493_2017}.  Briefly, the measurements have been performed on a diffractive optics based, all-reflective 2D spectrometer, with a phase stability of $\lambda/160$. The laser beam from a home-built nonlinear optical parametric amplifier (NOPA, pumped by a commercial femtosecond Pharos laser from Light Conversion) is compressed to $\sim$16 fs using the combination of a deformable mirror (OKO Technologies) and a prism pair. Frequency-resolved optical grating (FROG) measurement is used to characterize the temporal profile of the compressed beam and the obtained FROG traces are evaluated using a commercial program FROG3 (Femtosecond Technologies). A broadband spectrum so obtained carried a linewidth of $\sim$100 nm (FWHM) centered at 750nm.  Three pulses are focused on the sample with the spot size of $\sim$100 $\mu$m and the photon echo signal is generated in the phase-matching direction. The photon-echo signals are collected using a Sciencetech spectrometer model 9055 which is coupled to a CCD linear array camera (Entwicklungsb{\"u}ro Stresing). The 2D spectra for each waiting time T were collected by scanning the delay time $ \tau = t_{1}-t_{2}$ in the range of [-350 fs, 200 fs] with a delay time step of 2 fs. At each delay step, 100 spectra were averaged to reduce the noise. The waiting time $ T = t_{3}-t_{2}$ was linearly scanned in the range of 2.0 ps in steps of 15 fs. For all measurements, the energy of the excitation pulse is attenuated to 8 nJ with 1 kHz repetition rates. Phasing of the obtained 2D spectra was performed using an ``invariance theorem" \cite{JCP_115_6606_2001}.  

\subsection{Theoretical calculations.} 

A Frenkel-exciton model is constructed to calculate the coherent energy transfer and the 2D electronic spectra of the FMO complex. The total Hamiltonian is constructed in the form of the system, bath and system-bath interaction terms, $H = H_{S}+H_{B}+H_{SB}$. The system Hamiltonian is given as $H_S=\epsilon_g|g\rangle\langle{g}|+\sum_m^N\epsilon_m|m\rangle\langle{m}|+\sum_{m\neq{l}}^NJ_{ml}|m\rangle\langle{l}|$, where $\epsilon_g$ and $\epsilon_{m}$ are the site energies of ground and $m$th excited pigment, respectively. $J_{ml}$ is the electronic interaction between the $m$th and $l$th pigments. $N=7$ pigments are included in the monomeric FMO complex. Moreover, each pigment is coupled to its own individual bath.  The bath Hamiltonian can be written as $H_B=\sum_m^N\sum_{\xi}\hbar\omega_{m\xi}b_{m\xi}^{\dagger}b_{m\xi}$, where, $b_{m\xi}^{\dagger}(b_{m\xi})$ is the creation (annihilation) operator of $\xi$th fluctuation mode associated with site $m$, and its angular frequency $\omega_{m\xi}$. $H_{SB}=\sum_mV_mW_m$ describes the interaction of the system with the bath, where, we have defined $V_m=|m\rangle\langle{m}|$ and $W_m=-\sum_{\xi}c_{m\xi}(b_{m\xi}^{\dagger}+b_{m\xi})$. The $c_{m\xi}$ is the coupling constant between the $m$th pigment and $\xi$th fluctuation mode. The bath is specified by the spectral density $J_m(\omega)=\pi\sum_{\xi}\hbar^2c_{m\xi}^2\delta(\omega-\omega_{m\xi})$. We include one overdamped mode and one underdamped mode to study the impact of vibrational coherence. The corresponding spectral density can be expressed as $J(\omega)=\frac{2\Lambda\Gamma\omega}{\omega^2+\Gamma^2}+\frac{4S\gamma_{\mathrm{vib}}\omega_{\mathrm{vib}}^3\omega}{\left(\omega^2-\omega_{\mathrm{vib}}^2\right)^2+4\gamma_{\mathrm{vib}}^2\omega^2}$. Here, $\Lambda$ and $\Gamma^{-1}$ are the damping strength and the bath relaxation time of the overdamped mode, respectively. $S$, $\omega_{\mathrm{vib}}$ and $\gamma_{\mathrm{vib}}^{-1}$ are the Huang-Rhys factor, the vibrational frequency and the vibrational relaxation time of the underdamped mode, respectively.  This form has been shown to describe the experimental data \cite{PNAS_114_8493_2017} correctly. 

The nonequilibrium dynamics of the system-bath model is calculated by a time-nonlocal quantum master equation. The details of this method are described in the SI. Correlation theory is used to calculate the absorption spectrum of the FMO complex, $I(\omega)=\left\langle\int_0^{\infty}dte^{i\omega{t}}\mathrm{tr}(\bm{\mu}(t)\bm{\mu}(0)\rho_g)\right\rangle_{\mathrm{rot}}$, where $\rho_g=|g\rangle\langle{g}|$ and a $\delta$-shaped laser pulse is assumed.  $\langle\cdot\rangle_{\mathrm{rot}}$ denotes the rotational average of the molecules with respect to the laser direction. Moreover, the 2D electronic spectra are obtained by calculating the third-order response function 
\begin{equation}
S^{(3)}(t,T,\tau)=\left(\frac{i}{\hbar}\right)^3\Theta(t)\Theta(T)\Theta(\tau)\mathrm{tr}\left(\bm{\mu}(t+T+\tau)\left[\bm{\mu}(T+\tau),\left[\bm{\mu}(\tau),\left[\bm{\mathrm{\mu}}(0),\rho_g\right]\right]\right]\right). 
\end{equation}
Here, $\tau$ is the delay time between the second and the first pulse, $T$ (the so-called waiting time) is the delay time between the third and the second pulse, and $t$ is the detection time. To evaluate 2D electronic spectra, we need the rephasing (RP) and non-rephasing (NR) contributions of the third-order response function, i.e., $S^{(3)}(t,T,\tau)=S_{\mathrm{RP}}^{(3)}(t,T,\tau)+S_{\mathrm{NR}}^{(3)}(t,T,\tau)$. Assuming the impulsive limit (the $\delta$-shaped laser pulse), one obtains 
\begin{eqnarray}
I_{\mathrm{RP}}(\omega_t,T,\omega_{\tau})&=&\int_{-\infty}^{\infty}d\tau\int_{-\infty}^{\infty}dt e^{i\omega_{t}t-i\omega_{\tau}\tau}S_{\mathrm{RP}}^{(3)}(t,T,\tau), \\
I_{\mathrm{NR}}(\omega_{t},T,\omega_{\tau})&=&\int_{-\infty}^{\infty}d\tau\int_{-\infty}^{\infty}dt e^{i\omega_{t}t+i\omega_{\tau}\tau}S_{\mathrm{NR}}^{(3)}(t,T,\tau).
\end{eqnarray} 
The total 2D signal is the sum of the two, i.e., $I(\omega_{t},T,\omega_{\tau})=I_{\mathrm{RP}}(\omega_{t},T,\omega_{\tau})+I_{\mathrm{NR}}(\omega_t,T,\omega_{\tau})$. 

The model parameters of the site energies and electronic couplings are initially taken from Ref.\ \cite{JPCL_7_1653_2016} and further refined during a simultaneous fit to the absorption spectra of the FMO complex at the considered temperature. To precisely determine the reorganization energy, the parameters are further refined by fitting of the theoretical results to the experimental anti-diagonal bandwidth of the main peak of exciton 1.

%%%%%%%%%%%%%% 
%
\begin{addendum}
 \item This work was supported by the Max Planck Society and the Excellence Cluster  'CUI: Advanced Imaging of Matter' of the Deutsche Forschungsgemeinschaft (DFG) - EXC 2056 - project ID 390715994.  K. Ashraf and R. Cogdell acknowledge funding by ``The Photosynthetic Antenna Research Center'' under the  US DoE Energy Frontier Research Center grant number DE-SC 0001035. 

\item[Supporting information] The Supplementary Information includes the 2D electronic spectra measured at the temperatures 50, 80 and 150 K and the description of the global fitting approach and of the fitting procedures to retrieve electronic and vibrational coherences. The calculated 2D electronic spectra are presented from which the lifetimes of the electronic quantum coherences of higher exciton states are extracted. 

\item[Competing Interests] The authors declare that they have no competing financial interests.

\item[Correspondence] Correspondence of theoretical part should be addressed to M.T.  (email: \\ michael.thorwart@physik.uni-hamburg.de), Correspondence and requests for experimental details should be addressed to  R.J.D.M.~(email: dmiller@lphys.chem.utoronto.ca). 
\end{addendum}
%
%%%%%%%%%%%%%%%%%%%%%%%%%%%%%%%%%%%%%%%%%%%%%%%%%%
%%%%%%%%%%%%%%%%%%%%%%%%%%%%%%%%%%%%%%%%%%%%%%%%%% 

\newpage
%%%%%%%%%%%%%%%%%%%%%%%%%%%%%%%%%%%%%%%%%%%%%%%%%% 
%%%%%%%%%%%%%%%%%%%%%%%%%%%%%%%%%%%%%%%%%%%%%%%%%%

\newpage
%%%%%%%%%%%%%%%%%%%%%%%%%%%%%%%%%%%%%%%%%%%%%%%%%% 
%%%%%%%%%%%%%%%%%%%%%%%%%%%%%%%%%%%%%%%%%%%%%%%%%% 

\newpage

%%%%%%%%%%%%%%%%%%%%%%%%%%%%%%%%%%%%%%%%%%%%%%%%%% 
%%%%%%%%%%%%%%%%%%%%%%%%%%%%%%%%%%%%%%%%%%%%%%%%%% 

\newpage
%%%%%%%%%%%%%%%%%%%%%%%%%%%%%%%%%%%%%%%%%%%%%%%%%% 
%%%%%%%%%%%%%%%%%%%%%%%%%%%%%%%%%%%%%%%%%%%%%%%%%% 

\end{document}